\documentstyle[12pt,epsf]{article}
\newcommand{\bc}{\begin{center}}
\newcommand{\ec}{\end{center}}
\newcommand{\bd}{\begin{displaymath}}
\newcommand{\ed}{\end{displaymath}}
\newcommand{\be}{\begin{equation}}
\newcommand{\ee}{\end{equation}}
\newcommand{\ba}{\begin{array}}
\newcommand{\ea}{\end{array}}
\newcommand{\bea}{\begin{eqnarray}}
\newcommand{\eea}{\end{eqnarray}}
\newcommand{\bt}{\begin{tabular}}
\newcommand{\et}{\end{tabular}}

\newcommand{\bp}{\begin{picture}}
\newcommand{\ep}{\end{picture}}
\newcommand{\bfi}{\begin{figure}}
\newcommand{\efi}{\end{figure}}
\begin{document}

%\begin{figure}[t]
%\epsfig{file=PoSlogo.ps,width=\textwidth}
%\end{figure}

\title{{\huge \bf Complex Action Support from Coincidences of Couplings }}

\author{ 
H.B.~Nielsen 
%${}^{1}$ 
\footnote{\large\, hbech@nbi.dk} 
%\\[5mm] 
\footnote{This article corresponds to a talk at the Bled Conference 
on ``What comes beyond the Standard Models'' organized by Norma Mankoc 
Borstnik et al. but were not put in the proceedings; but the talk 
is to be found on Cosmovia on which it were transmitted in July 2010} \\[5mm] 
\itshape{
%${}^{1}$ 
The Niels Bohr Institute, Copenhagen, Denmark}}

\date{}

\maketitle

\begin{abstract}
Our model  \cite{ownmMPP} \cite{SIMPP}with a complex action in a 
functional integral formulation with path integrals extending over all
times, both past and future, is reviewed. Several numerical relations between 
coupling constants are presented as supporting evidence.The new evidence 
is that several more hitherto unexplained coincidences are explained by our 
model:

1) The ``scale problem'' is solved because the Higgs field expectation 
value is predicted to be very small compared to say some fundamental scale,
that might be the Planck scale.

2) The Higgs VEV need not be just zero, but rather is predicted to be so that 
the running top-quark Yukawa coupling just is about to e unity at this scale;
in this way the (weak) scale easily becomes ``exponentially small.
Instead of the top-Yukawa we should rather say the highest flavour Yukawa 
coupling here.

These predictions are only achieved by allowing the principle of minimization 
of the imaginary part of the action $S_I(history)$ to to a certain extent 
  adjust some coupling constants in addition to the initial conditions.

If supersymmetric partners are not found at LHC it would strengthen the need 
for a ``solution'' of the hierarchy problem in our direction of an explanation 
via a finetuning scheme inside the Standard Model, from say minizing ``the
imaginary part of the action'' in our complex action model.  
\end{abstract}

\newpage
\thispagestyle{empty}
\section{Introduction}
Masao Ninomiya and myself \cite{ownmMPP} \cite{SIMPP} have since some time 
worked on the idea 
that the action going into the Wentzel-Dirac-Feynman-path-integral
\cite{FH}\cite{Wentzel}\cite{Dirac}
should be {\em complex}, rather than as normally assumed, real. Alone to argue 
that such a model would not lead to so drastic predictions deviating 
from usual observations, that it is not falsified immediately, needs a 
highly nonrivial speculative argumentation. One step is to first see 
that the effect of the imaginary part of the action first of all is
{\em determine the initial conditions} by selecting which solution 
to the equations of motion is to be realized, and that according to 
a minmization rule of this imaginary part of the action $S_I(history)$,
a quatity depending a priori on what goes on at all times. 
Then, however, we can suggest   that the time close to the 
effective Big Bang time would be much more important for determining 
the development, so that not much freedom in ``choosing'' it would be left 
to cause effects of the imaginary part to be seen at later times, is not 
unnatural in our model.
%However, alone the fact, 
%that in a natural way such a complex action theory could produce so small 
%deviations from the usual real action theory that nobody would have 
%discovered it untill now, is a highly interesting point, because it allows 
%for that the action could indeed be complex. 
To have a complex action  would in fact  be simpler 
in the 
sense
that as far as the integrand $ \exp(\frac{i}{\hbar} S(path)) $ of the 
path way integral is anyway complex,  whether the action $S(path)$ is allowed 
to be complex or only to be real. Then one could namely take the 
philosophy, that the most ``fundamental'' object were the {\em integrand} 
rather than the action and would even in that philosophy not have to impose 
any strange extra assumption on this integrand such as the usual one that it 
be a pure phase factor. These type of ideas leading 
to influence comming from the future too were indeed 
forerun \cite{old}\cite{vacuumbomb} by works by  Don Bennett and myself  and 
by bayuniverse theory \cite{babyBanks}\cite{babyColeman}\cite{babyHawkings}
\cite{HLbaby}   \cite{selfbaby}.  

It is the main purpose of the present article to put forward a couple of 
numerical coincidences that support the complex action model in the sense 
that they can be derived form the latter under mild and natural assumptions 
within it and which seem to agree with the experimentally measured parameters.

One such relation I already published in \cite{Spaatind} and that means that 
there is relation between the light quark masses and the electron and 
nucleon masses in a way that would be hard to understand as more than 
accidental in any other model. Also in that article were mentioned 
the funny accident that the ``knee''\cite{Knee} in the cosmic ray spectrum
\cite{Gaisser} 
- an energy at which the intensity of cosmic ray impacts seeming starts 
falling faster with energy than below - happens to be very close in order 
of magnitude to the effective threshold for significant Higgs boson 
production.

The main new point of the present article is, however, what we could consider 
a solution of the scale problem - related to the hierarchy problem of why 
the weak energy scale (or say Higgs mass scale) is so enourmously low compared 
to say the Plack scale, or if there were some GUT then compared to the 
GUT-scale. 

The main point really is that we argue in the development of the complex 
action model for that the {\em square} of the Higgs boson field averaged 
is the quantity that should be minimized so as to fix say the bare Higgs 
mass square and thereby the Higgs field expectation value $<\phi_h(x)>$
to minimize the square.
It is intuititively obvious that minimizing the {\em square} leads 
easily to make the average of the Higgs field itself become small.
We shall see, however, that nearer estimates involving the influence of 
virtual top-quarks in the vacuum very likely leads to a correction that 
potentially could make it non-profitable in searching for the lowest possible 
average of the {\em square} of the Higs field $<\phi_h(x)^2>$ to just decrease 
the expectation value of the Higgs field itself unlimited towards  zero. Rather, 
when the 
top-Yukawa coupling gets sufficiently large, it could and we shall suggest 
that it does go oppositely, and it thus pays to stop decreasing the Higgs 
field 
expectation value $<\phi_h(x)>$ at the situation wherein the running 
top-Yukawa coupling at the scale of this $<\phi_h(x)>$ passes a certain value 
of 
order unity.   

Since the running of say the top Yukawa coupling $ g_t(\mu)$ is 
``logarithmically slow'' the value of an energy scale determined 
by putting such a running coupling to a special value, such as 
one of order unity, can easliy lead to an ``exponetially small'' energy scale.
It is in this way our work points to a possible explanation for how the 
weak scale related to say the Higgs mass or the Higgs field expectation 
value  could become naturally ``exponetially small''. That is to say 
our model based on the complex action leads to a solution of what we may call 
the ``scale problem'' of ``Why is the Higgs or weak scale  so extremely 
low compared to Planck scale?''. Having such an explanation of an 
{\em adjustment mechanism} making the Higgs mass say small one should 
of course in principle use it order by order in perturbation theory
in the Standard Model. Thus although there would still be a problem 
of tuning in the bare Higgs mass square  by huge terms compared to 
the renormalized squared mass, we would in our model have an explanation 
for why we have to
renormalize in this ``fine tuning'' way and thus in a certain philosophical 
sense we would have ``solved the hierarchy problem'', in spite of there still 
being  huge renormalizations (quadratically divergent contributions in 
the bare 
mass square).

Since we shall claim to explain the smallness compared to the Planck scale 
of the Higgs field vacuum expectation value - as what minimizes the Higgs 
field 
squared and averaged-, we can claim that we propose a picture for which there 
would be a stronger call if the susy-partners are not found at LHC. Then 
namely we cannot even solve the hierarchy problem by susy, and it will be 
really needing with some finetuning law rather than avoiding finetuning 
totally. We have got the minimization of the $S_I$ to lead to finetuning, that may 
thus help solving even hierarchyproblem at the end.  

After in section \ref{review1} and section  \ref{review2} to have reviewed 
very shortly the complex action model by  Ninomya and myself  \cite{ownmMPP}
\cite{SIMPP} with 
special emphasise in section \ref{review2} on the approximation that the 
Higgs mass term is supposedly dominant, we shall in section \ref{couplings} 
introduce the assumption that also some coupling constants shall  be adjusted 
to minimize the imaginary part of the action $S_I$. Next in  section 
\ref{phi2verphi} we
discuss how we expect the average of the square of the Higgs field to vary 
as a function of the average of the Higgs field itself, when we vary say the 
(real part of the) bare Higgs mass square $m_h^2|_R$, i.e.the coefficient of 
the Higgs field  
square term in the Lagrangian density of the Standard Model.
It is in this section \ref{phi2verphi} we achieve the main result of 
predicting the Higgs expectation value from minimizing the average of the 
square of the Higgs field and predict the scale to be the one where the 
top-Yukawa coupling 
is of order unity.   
Then in section 
\ref{toomuchque} we shall discuss whether a model with a fixing of 
initial conditions as well as coupling constants to some extend   
like our model will not lead to some strange effect, such as cleaning the 
universe for neutrons or oppositely making all the proton and electrons into 
neutrons. This worry turns out to well worth worrying about. The point, 
however, turns out to be that there is just such a relation between certain 
coupling constants - the light quark masses the elctron mass and the 
energy of the valence quarks in the nucleon - so that it does not matter if we 
let the neutron decay or not for evaluating the imaginary part of the 
action $S_I$. This formula is reviewed from reference \cite{Spaatind} 
in section \ref{relation}. 
Before concluding we put forward a couple of other adjustments to make our 
minimizing Higgs field square integrated principle consistent in section
\ref{further}. Especially it is new that I here in subsection \ref{binding}
seek to show that our model only shall be consistent provided 
the binding of nucleons into nuclei is remarkably weak, as we would 
say it is.
Finaly in section \ref{conclusion}
we shall conclude that there has actually collected itself a few 
numerical coincidences strongly suggesting that there ought to be 
something about the complex action model, since otherwise why should 
they be satisfied even approximately?
     
\section{Short review of complex action model}\label{review1}
Looking at the Wentzel Dirac Feynman path integral \cite{FH,Wentzel,Dirac} 
\begin{equation}
\int \exp (\frac{i}{\hbar} *( S_R[path] +iS_I[path])){\cal D } path
\end{equation}
for the case that the action is considered being complex
\begin{equation}
S[path] = S_R[path] + i S_I[path] \label{action}
\end{equation}
(where of course $S_R$ and $S_I$ just are the real and imaginary parts of the 
action  $S$)
it is clear that almost whatever the interpretation procedure for the use 
of this path-integral may be taken to be it will be so that only the 
paths for which 
the imaginary part $S_I[path]$ is close to being minimal - at least 
among those paths, which corresponds to some classical solution - 
can contribute significantly. In the light of that having the $\hbar$
in the denominator suggests the coefficient $\frac{i}{\hbar}$ to be in 
some ``human terms''  very large we would even expect that probably only 
the very minimum value of the imaginary part of the action $S_I$ will
be realized for any path of significance. That fact we take to mean physically 
that the history of the universe which is  actualy realized - the one we 
live through - thinking in a classical approximation must be that history
among the ones agreeing with the equations of motion at all that has the 
minimal value for the imaginary part $S_I[history]$. 
This is a first remarkable result of the investigation of the complex action 
model, that it becomes {\em a theory of the initial conditions} and not only 
for the equations of motion as usual theories have it.

Next remark that since in the way we did it with inclusion of both past, 
present and future into the time-integral defining the action (\ref{action}) 
\begin{equation}
S[path] = S_R[path] +i S[path] = \int_{-\infty}^{\infty}L(t)dt
\end{equation}
where of course now $L(t) = L_R + iL_I(t)$ is the complex Lagrangian,
the selection of the {\em initial conditions and thereby solution being 
selected to be the realized one by having minimal $S_I$ now depends on the 
future too.}. This indeed means as a very remarkable property of theories 
of this type that it will be as if there is some arrangement going on in the 
world achieving certain happenings in the future. That is to say that our 
complex action model has in it {\em backward causation}. If for instance 
as we earlier suggested Higgs particle production in big amounts 
(in \cite{Spaatind} the number of order 300000 Higgses is mentioned 
as possible critical number in the sense that more Higgses 
in a single machine should presumably be avoided, if it is at all possible 
by adjustment of the initial conditions) should be avoided by such 
prearrangement of the historical development of the Universe. 
There  should essentially appear a bit of miraculously bad lucks for Higgs 
producing 
machines (making more than the critical number order of magnitude),
so that they some way or another by bad luck turn out not to come to function.

There has been in the development of the understanding of the consequences 
of our complex action model some difficulty in speculating out how 
to avoid that these effects of backward causation or prearrangements, 
avoiding e.g. Higgs production, become so big that there would be no chanse 
that they should not have been observed. If indeed our complex action model
predicted that there would be several happenings that would be avoided 
while other types of events would/should show up more copiously than we
would expect from statistics  
%be so commonly observed that it would mean that 
our model would be  killed already from the very fact that we do not 
see much of 
such effects in dayly life. 
If we should make an a priori guess as to the size of the imaginary part 
$S_I$ of the action it would be 
%Really the most likely and nice guess would be
that the real part $S_R$ and the imaginary part $S_I$ would have 
coefficients to the various terms in the Lagrangian density being of the same 
order of magnitude. We hope that we can still uphold such a crude 
hypotesis of the coefficients in the Lagrangian density, which  would mean 
that the coefficients (coupling constants, mass squares) would just be 
complex numbers with phases of order unity measured in radian. But  such  a
situation would in  
very naive thinking lead to the a priori exponential supression factors 
of the the integrand in the path way integral, and that with a very big 
exponent, which even has the expected tremendously big factor $1/\hbar$
in front. Really it would mean that the exponential suppression would 
be by the exponent of a number of the order of 
 the phase variation of the waves in quantum mechanics for 
the system in question. That would be enourmously much and the observation 
of the effects of backward causation could hardly be avoided in dayly life.
We hope to have developped at least some - may be a bit speculative - arguments
that it is not excluded that there are very strong suppressions of the 
backward causation effects relative to this just mentioned naive estimation. 
The main argument for strong suppression of the effects of backward causation 
from the principle of minimizing $S_I[history]$ is that supposing that we 
have to have the equations of motion fullfilled what happens at one moment of 
time and at another moment of time gets related simply by integrating the 
equations of motion. If therefore something shall be arranged at one moment 
of time it leaves less freedom of adjustment of the initial conditions 
for the instalment of special happenings at another moment of time. Since 
the Universe already has existed for a very long time compared to the 
periode of
time about which we have sufficient knowledge to be able to discover effects 
of backward causation or prearrangements the strength of the backward 
causation of which we can hope to be aware gets already by this consideration 
of many competing eras strongly suppressed. In order to have agreement 
with the fact of increasing entropy we have to speculate that the eras
of most importance for deciding the history with the minimal $S_I$, i.e. 
the realized history, were the ones just after a possible big bang
or more precisely the eras close to the very special and hot conditions etc 
arround the time of which we usually think as the big bang time. This 
hypotesis that 
the eras near the special or big bang time dominate the selection of the 
history is not at all unlikely inside our complex action model, because 
at that time fields such as inflaton field or plasmas had very high energy 
densities compared to the present time. That means then that a Lagrangian 
density having the same dimension as the energy density would very likely 
get enourmously much bigger constributions during such a hot era than 
under the later cold eras even if there would be a compensation due to the
Universe being bigger and the time spans large later on. In fact the energy 
density and thus presumably also the Lagrangian density - real as well as 
imaginary parts - goes in say radiation dominated universe as 
\begin{equation}
{\cal L}(x) \propto 1/a^4
\end{equation}  
 where $a$ is the size/radius of the universe. If we thus 
take for simplicity that the time span relevant is of order $a$ like
the spatial size, the integral to evaluate the action - real or 
imaginary part - becomes of the crude form
\begin{equation}
\int \frac{1}{a^4}d^4x \approx \int \frac{da}{a}.
\end{equation}
This means that it gets about similar order of magnitude contributions from 
each order of magnitude of $a$ or in our crude thinking here from each order 
magnitude in the time span. But if is it so there is a logarithmic divergence 
towards the big bang. This was very crude but it means that there could 
exceedingly easily be some of the time intervals from us to see very close to 
the big bang which could dominate.

\section{The special importance of the Higgs field squared}\label{review2}

In our times when it is the fields of the Standard Model that seemingly 
plays the most important role a concrete guessing for the action is 
of course the Standard Model action 
\begin{equation}
\int {\cal L}(x) d^4x
\end{equation}
where the Lagrangian density ${\cal L}(x)$ has several terms of which we 
just mention a few to illustrate
\begin{equation}
{\cal L}(x) = -\frac{1}{4 g_s^2} F^a_{\mu\nu}(x)F^{a\mu\nu}(x) + ...
+ |D\phi_H|^2 - m_h^2 |\phi_H|^2 + ...
\end{equation}
%is the most important. 
It would therefore be the most natural suggestion 
that in our times the imaginary part of the action $S_I$ simply gets 
its main contributions from terms of precisely the same form just that 
it is now the imaginary part of the coefficients such as 
$\frac{1}{4g_s^2}$ and $m_h^2$  i.e. 
 $\frac{1}{4g_s^2}|_I$ and $m_h^2|_I$ which gives us the imaginary part 
of the action 
\begin{equation}
S_I[path] = \int {\cal L}|_I(x) d^4
\end{equation}
where now
\begin{equation}
{\cal L}(x) = -\frac{1}{4 g_s^2}|_I F^a_{\mu\nu}(x)F^{a\mu\nu}(x) + ...
+Z_I |D\phi_H|^2 - m_h^2|_I |\phi_H|^2 + ....
\end{equation}
(On the kinetic term, which is often normalized to have unit coefficient, 
we had of course to insert for the case of the imaginary part a parameter 
or coefficient $Z_I$, since we cannot normalize the field - in the example the 
Higgs field $\phi_H(x)$ to the imaginary part, also.).

As already suggested in foregoing section it would be natural to expect 
the real and imaginary parts of the coupling coefficients to be of similar 
order of magnitude. There is, however, one case in which this may not 
be trustable, if we thought of the renormalized coefficients, and that is the 
case of the Higgs mass term. The renormalized Higgs mass is namely surpizingly 
small in the sense of the scale problem being the problem of why it is so  
very small compared 
to, say, the Planck mass. Remember it is a kind of mystery that it is so small,
a mystery needing one explanation or an  other (we shall actally give an 
explanantion attempt ourselves in the next section) and thus it is far 
from obvious whether we should believe that the same mysterious mechanism 
making the renormalized real part of the mass square of the Higgs small 
should also work for the imaginary part. But most importantly we should 
guess that it is the bare Higgs mass coefficient that has the about equally 
big real and imaginary part rather than for the renormalized or dressed 
mass square. Now truly the hierarchy problem is about that we can not 
get the bare mass square and the renormalized mass square ``small'' - meaning 
of the order of the weak scale - simultaneously unless we have invented 
some way (such as susy) to make the renormalization correction ``small''.
But it is only the renormalized real part that we know is ``small'' so
unless we have the very clever mechanism making the renormalization correction 
``small'' we expect that the imaginary part of the mass square is not 
``small'' at all. The most suggestive expectation is indeed that the 
imaginary part of the mass square of the Higgs is say Planck scale big, 
renormalized or not. But that then makes the term 
$...+m_H^2|_I |\phi(x)|^2 +...$ in the imaginary part of the Lagrangian 
density become enourmous from the point of view of the $100 GeV$ scale say.
This not believing in any``mysterious'' mechanism making the imaginary part 
$m_H^2|_I$ small means that it becomes about $10^{34}$ times bigger than the 
other terms for energies of the $TeV$ or so.

This means that under the experimental conditions of dayly life and 
realistic accelerators and even essentially all astrophysics of today 
presumably not even excepting cosmic rays we must count that the Higgs mass 
(square) term in the Lagraingian density completely dominates w.r.t the 
imaginary part. We can therefore except only for perhaps times 
deep inside the first secund in cosmology use the approximation for the 
imaginary part of the action
\begin{equation}
S_I[path] = - m_h^2|_I \int |\phi_H(x)|^2 d^4x. \label{mainap} 
\end{equation}   

Since the universe has not been filled with huge amounts of Higgses we shall
take it that the sign of the coefficient $m_h^2|_I$ is so that the term 
gets positive for big Higgs field, so that the minimization of 
$S_I$ resulting from our model would tend to make there be only few Higgses.
That is to say that in the skechted notation (\ref{mainap})
\begin{equation}
m_h^2|_I < 0.
\end{equation}

Now most of the Universe is today practically vacuum or at least very 
close to being just empty space. In the vacuum, however, there is a 
fluctuating Higgs field $\phi_H(x)$ - that does as we know not even 
fluctuate arround zero, but rather arround a ``small'' value 
$246 \ GeV/\sqrt{2}$ in the sense of smallness just described
(i.e. compared to Planck units). 
Because of the enourmous spacewise dominance of the vacuum today
we expect the most important term in the imaginary action to be minimized 
to be the approximation:
\begin{equation}
S_I[path] \approx -m_h^2|_I <|\phi_H(x)|^2 > \int d^4x 
\approx -m_h^2|_I <|\phi_H(x)|^2> VT
\end{equation}       
where $VT$ is the four volume of space-time (of course $V$ is the volume 
and $T$ the time of duration of the world.).

\section{Let also the $S_I$-minmization Influence the Couplings}
\label{couplings}
In principle the expectation value $<|\phi_H(x)|^2>$ in vacuum is given 
alone by the coupling constants of the theory, except for the case that 
there are several vacua, and thus it looks at first that there is no way to 
choose the {\em initial state conditions} so as to arrange anything 
for say this average of the square of the Higgs field $<|\phi_H(x)|^2>$
to be small/minimal. There are though many ways in which such a dependence 
on the initial state conditiond for even the vacuum can {\em ``sneak in''}
by some effective coupling constants depending on initial conditions:

It happens in baby universe theory, or if there were indeed very many possible 
vacua like in the landscape story by Caroll, Smolin, Susskind 
\cite{landscape}, or if just the 
fundamental physics were somehow say a pregeometry with many more 
degrees of freedom than the ones we know so far. There could easily be some 
degrees of freedom, that could somehow set the vacuum to have various 
properties(such as say a small Higgs field square average).In string theories 
there is a lot of plays with ``moduli'', which are degrees of freedom that 
function as coupling constants. Such moduli would clearly - being considered 
as part of initial conditions and thus adjustable in our model to minmize 
$S_I$- be able to adjust the Higgs field expectation value.. By nature 
adjusting some such moduli according to the minization 
of $S_I$ it would namely also 
adjust the Higgs field expectation value, say, since that would be a function 
of the moduli. So in such 
a way the effects of adjusting to minimize $L_I$ or $S_I$ could easiliy 
``sneak in'' via some sort of moduli. 

If we subscribe to one of these kinds of possibilities for the vacuum having 
some yet to be understood degrees of freedom behind it, it would mean that 
we should  for the purpose of minimizing $S_I$ consider at least 
some of coupling the constants or better some combinations of them or 
parameters 
parametrizing them as (dependend on) initial conditions. That should then 
simply mean that at least still keeping some relations between the couplings
perhaps there would be some freedom in varying the couplings, that then 
in our model should be adjusted by minimizing the imaginary part of the 
action in the same way and together with the genuine initial conditions.

That is to say that we could for instance imagine that the {\em real 
part of the Higgs mass square $m_h^2|_R$} might be adjustable and should 
be determined so as to minimize $S_I$. That would in our approximation
(\ref{mainap}) mean that the real part of the Higgs mass square should be 
adjusted so as to minimize the average of the square of the Higgs field 
$<|\phi_H(x)|^2>$. 

So our model predicts that for instance the bare Higgs mass square coefficient 
gets tuned in to make the avegrage of the squared Higgs field $<|\phi_H(x)|^2>$
minimal. It does not really matter so much whether it is just the 
Higgs mass or something related, but at first we have intuitively that it 
should be very favorable for making  $<|\phi_H(x)|^2>$ small to make the 
Higs field vacuum expectation value $<\phi_H(x)>$ (without squaring)
small or zero. So we could also for instance think of the expectation value 
as the quantity that gets adjusted, since it should not matter much what we 
think of as what we can screw on.

\section{The Squared Higgs field expectation value  as function of the 
Higgs field VEV}
\label{phi2verphi}

\subsection{Crudest approximation}
How do we now expect the square of the Higgs field averaged in vacuum  
$<|\phi_H(x)|^2>$
to depend on the average of the average of the Higgs field itself 
$<\phi_H(x)>$ without the square in the range of rather small values in which 
 we are most interested ?
%From the point of view of thinking on low values as we would think in 
%the units of present accelerators 
The first estimate is simply to ignore quantum fluctuations and say that 
of course we simply have ignoring fluctuations
\begin{equation}
 <|\phi_H(x)|^2> \approx |<\phi_H(x)>|^2 \hbox{(ignoring fluctuations)}.
\label{cl}
\end{equation}
If this approximation were true, one would predict from our model 
of course, that the expectation value of the Higgs feild in vacuum 
$<\phi_H(x)>$ should be zero. Note that in the light of thinking on the 
Planck units as the fundamental units in nature this prediction is indeed 
already a very good prediction answering to the main problem of 
why the Higgs expectation value is so enourmously small compared to the 
Planck scale expectation. However we know from the whole application and 
reason for our belief in the existence of the Higgs field at all that the 
Higgs field vacuum expectation value shall not be zero but just very small 
compared to the Planck unit. Before going to a more accurate calculation 
of the Higgs field vacuum expectation value (VEV), let us shortly 
remember that 
really we should rather than the VEV $<\phi_h(x)>$  predict some parameter 
in the Lagrangian density, which really here means the real part of the 
mass square of the Higgs $m_h^2|_R$. Now you can obtain in the classical 
approximation (\ref{cl}) the zero VEV for all positive $m_h^2|_R$. It would 
first be in some further degree of accuracy that we would get the value 
the real mass square $m_h^2|_R$ giving the minimal $<\phi_h(x)^2>$. Since 
classically this average of the square   $<\phi_h(x)^2>$ goes as a constant 
for $m_h^2|_R > 0$, i.e. non-tachyonic case, any very small $S_I$ contribution 
could very easily drive it to one end or the other of the interval giving the 
minimum. Thus it would have a high probability that for instance the mass 
square $m_h^2|_R$ would be driven to just zero. In this sense already in the 
classical approximation (\ref{cl}) we have argued not a finetuning mystery, 
but rather a likely consequence in our model to obtain
\begin{equation}
{\bf Classically : } \quad  \quad  m_h^2|_R \approx 0 \quad  %\hspace{(``classically'')} 
\end{equation}

\subsection{Including fluctuations}

Now my claim is that we have to make a bit more accurate estimate of the 
Higgs field squared expectation value $<|\phi_H(x)|^2>$ as a function of the 
vacuum expectation value $<\phi_H(x)>$ without the square. There may be 
several corrections since the fluctuation of the Higgs field 
\begin{equation}
``fluctuation'' = <|\phi_H(x)|^2> - <\phi_H(x)>^2 
= <|\phi_H(x) - <\phi_H(x)>|^2>
\end{equation}   
does not have to be  independent of the parameters such as $m_h^2|_R$ which 
is used to let the expectation value $<\phi_H(x)>$ vary. Realisticly 
the fluctuations will actually be huge compared to the classical contribution 
$|<\phi_H(x)>|^2$ to the average square $<|\phi_H(x)|^2>$ in the situation 
known phenomenologically of a surprisingly small Higgs field average.
The size of the fluctuation will then mainly be given by the fourth order 
term $-\frac{ \lambda}{8} |\phi_H(x)|^4 $ in the Lagrangian density and thus 
be approximately independent of the mass- coefficient $m_h^2|_R$. Still the
mass term has some effect, but there is yet another type of effect which 
could be even more important in the region of very small VEV $<\phi_H(x)>$ 
and this  is the most intersting. That is the higher order effect of the 
the fluctuation contribution due to virtual top-qurks say in the vacuum.
Certainly there will be in the vacuum some top-quark anti-top-quark pairs 
virtually. It may be better to think of it by saying that the top-quarks in 
the Dirac sea are in quantum superpositions of having different positions and 
so there are quantum fluctuations in these positions at least. That means that 
there are top quarks here and there in a quantum random way. Around these 
top-quarks there will then be their Yukawa-Higgs  fields meaning that 
in the neighborhood of the in this way virtual top-quarks the normal Higgs 
field value will be a bit suppressed. Of course then compared to the average 
there will  to cope with the definition of the average be a bit enhanced 
Higgs field in the regions where there is accidentally relativly few virtual 
tops. Obviously this is expected to mean that the stronger 
the coupling of the top to the Higgs, i.e. the stronger the top-Yukawa-coupling
$g_t$ relevant for the scale considered, the more fluctuations there will be 
due to the virtual top-quark Yukawa fields. If the top-Yukawa coupling is 
truly so small that the effect of it is only a small correction it may not 
matter 
much for how the expectation of the square $<|\phi_H(x)|^2>$ behaves as 
function 
of the average $<\phi_H(x)>$; but if the Yukawa coupling becomes of the 
order of unity in the sense that the virtual top fluctuation contribution 
becomes of the same order as the contribution $|<\phi_H(x)>|^2$ via
the classical approximation to the average of the square   $<|\phi_H(x)|^2>$,
this virtual top contribution could be significant.
Actually one could imagine that once the running top-Yukawa $g_t(\mu)$ 
%(where $ t = \ln(\frac{\mu}{M_Z}$ is the scale logarithm representing
%quantization scale $\mu$) 
 has become of order unity, then contrary to classical expectation 
the average of the square  $<|\phi_H(x)|^2>$ would no longer be an increasing 
function of the expectation value $<\phi_H(x)>$ but rather decreasing in stead.
The point is that varying the expectation value $<\phi_H(x)>$ the top
quark mass increases with increasing expectation value and thus the 
density of virtual top quarks decrease, leading in turn to a decrease 
of the fluctuation in the Higgs field due to these virtual top quarks.
Thus if this effect of the virtual top quarks is sufficiently strong it 
might overcompensate for the classical effect w.r.t how the average square 
$<|\phi_H(x)|>$ varies as function of   the average $<\phi_H(x)>$. When these 
two 
compensating effects just balance there will be a minimum or a maximum 
in the average of the square $<|\phi_H(x)|^2 >$ as a function of the average 
itself $<\phi_H(x)>$ of the Higgs field. Whether it becomes a maximum or a 
minimum depends on the sign of the $\beta$-function for the top-Yukawa 
coupling at the relevant scale $\mu$. We shall indeed easily see by 
using the expression for the top-Yukawa beta function that in the weak 
scale range, which is of interest for us, it would if the coupling gets 
sufficiently strong to make either a minimum or a maximum.
% indeed be a minimum:
   
We shall see that it indeed becomes a minimum:
     We shall use  the perturbative result \cite{beta} 
\begin{equation}
\beta_t^{(1)} = -(\frac{17}{20}g_1^2 + \frac{9}{4} g_2^2 + 8g_3^2)g_t + 
\frac{9}{2}g_t^3 +\frac{3}{2}g_tg_b^2
\end{equation}
with SU(5)adjusted convention for the coupling constants, where 
we have defined
\begin{equation}
\frac{dg_y}{dt}= \bar{\beta_y}
\end{equation} 
with e.g.$y=t$ for the top quark Yukawa coupling $g_t$ for the top,
(for more than one loop we must include the anomalous dimension $\gamma$ and 
use 
\begin{equation}
\bar{\beta_x} = \frac{\beta_x}{1-\gamma} 
\end{equation}
and to one loop just the equation
\begin{equation}
\bar{\beta_x } = \frac{\beta_x^{(1)}}{(4\pi)^2}.
\end{equation}
Then we see that for the couplings in the weak scale range, where 
$g_t \approx 1$ and $g_3 \approx 1 $ too, while the $g_1$ and $g_2$ are 
smaller it is the two terms $ -8g_3^2g_t + \frac{9}{2}g_t^3 $ that are 
important. It is actually so that there is a ``quasi fixed point''\cite{qfp}
meaning that the terms in the beta-function roughly cancels and the running 
coupling almost became fixed. Once the qcd-coupling $g_3$ however would 
become big as the QCD-scale is approched the important negative term 
$ -8g_3^2g_t$ 
in $\beta_t^{(1)}$   will of course take over. Thus the running top Yukawa 
coupling $g_t$ will grow smaller as energy goes up inside  this range, and 
correspondingly of course it will grow stronger as the energy scale is lowered.

Our major point now is that if the energy scale $ \mu =<\phi_h>$ as given 
by the Higgs field 
expectation value $<\phi_h>$ is put very low, there will be effectively 
a very strong running Yukawa coupling $g_t(\mu)= g_t(<\phi_h>)$ for the 
virtual top-quarks 
that will be present in a vacuum state with such a very small energy scale.
This means that their Yukawa fields arround these virtual top-quarks will have 
a strength depending on the running Yukawa coupling and finally the Yukawa 
field 
contribution from the mentioned virtual tops in the vacuum will increase 
as the expectation value of the Higgs field is lowered. It can even at some 
point increase so fast under the scale lowering that it might {\em 
 overcompensate } for the decrease of the Higgs field square expectation 
value $<|\phi_h|^2>$ due to the decrease of the expectation value 
$<\phi_h>$ itself. To decrease this Higgs field expectation value 
$<\phi_h>$ further after that point would no longer pay w.r.t. diminishing 
the square $<|\phi_h|^2>$ or roughly equivalently the imaginary part
of the action $S_I$. Thus there will be a {\em non-zero value 
of the Higgs field expectation value $<\phi_h>$, which provide the minimal
value for the square $<|\phi_h|^2>$, and thus becomes the favourite 
value according to our minimizing the imaginary part of the action $S_I$
principle}. This means that our complex action model predicts a small but 
non-zero Higgs expectation value. In fact this small favourite value of the 
expectation value $<\phi_h>$ is characterized by being the one at 
which the changes with further change in the scale $<\phi_h>$ due to the 
Yukawa fields around the virtual top quarks in the vacuum and those
due to the direct change in the square $<|\phi_h|^2>$ due to the change 
in the average 
$<\phi_h>$  just compensate. 

\subsection{Crude estimation of the virtual top quark contribution to the 
Higgs field fluctuations}

Let us first have in mind that since we are interested in a 
{\em small/infinitesimal change } in the scale given by the average of 
the Higgs field it is only the field fluctuations,
which are changed when we change the scale i.e. change the expectation 
value  
%around that scale 
%i.e. around the expectation value 
$<\phi_h>$,
which matters. Thus we shall 
most crudely 
5e.g. only 
consider only the virtual top quarks 
having momenta in a range of this order in 
order to look 
%our estimate of looking 
for the 
balance point (where $<|\phi_h(x)|^2 >$ has minimum). To be definite we could 
thus consider instead of all 
top quarks in the Dirac sea only those with momenta in say a range 
in which the numerical value of the momentum varies by a factor $e= 2.71...$
around a midle value taken as the  value $<\phi_h>$. Such a range corresponds 
to a volume in 
three-momentum-space $\mu^3 *4 \pi$ because the surface of a sphere in 
three-space  is 4 $\pi$ times the radius $\mu$ squared. Now we know that 
in phase space with three degrees of freedom  the density of quantum states is
$1/h^3 = 1/(2\pi)^3$ (where we have put $\hbar =1$ so that Plancks constant 
without the bar is $h=2\pi$). So the density of Dirac sea Fermions 
in the momentum range described 
per unit 
space volume is $ \mu^3 *4 \pi /(2\pi)^3 = \frac{\mu^3}{2\pi^2}$. We should 
multiply this density by the number of color and spin states of the type of 
Fermion, we have in mind, here the top quark. For the top quark this becomes 
 a factor 6. Next we should estimate what fraction of a Fourier resolution 
of the Yukawa field around a virtual top quark taken here to mean one in the 
Dirac sea with momentum in the range shall be counted as in our chosen 
range of wavelengths. Sufficiently accurate for our anyway rather crude 
estimate we might take the part of the Yukawa potential around the (virtual/
or Dirac sea) top-quark lying inside say a distance $\sqrt{e}/\mu$ from 
the topquark position and outside $\frac{1}{\sqrt{e}\mu}$.
The idea by this cutting away of the least  steep and most steep parts of 
the potential 
is that the not so steep  part would contribute wave numbers no longer 
of order 
$\mu$ but rather smaller than $\mu$ and the most steep part 
would contribute mainly to bigger wave numbers than $\mu$. Let us remember 
that the top Yukawa 
field is 
\begin{equation}
V_{Yukawa} = \frac{g_t/2 }{4\pi* r}
\end{equation}
and so the integral over the square of it 
%from the distnce $r$
%(from the quark to the running point) being for simplicity $0$ to  
%$\sqrt{e}/\mu$ (it does not matter much in the small $r$ end, but 
%really we should take the region of 
$r$ being  $1/(\sqrt{e}\mu)$ to 
$r$ being $\sqrt{e}/\mu$ )
becomes
\begin{eqnarray}
\int_{1/(\sqrt{e}\mu)}^{\sqrt{e}/\mu} 4\pi r^2 * V_{Yukawa}^2 dr&=& 
(g_t/2)^2\int_{1/(\sqrt{e}\mu)}^{\sqrt{e}/\mu}\frac{dr}{4\pi}\\
& =& \frac{\sqrt{e}-1/\sqrt{e}}{4\pi * \mu}. 
\end{eqnarray} 

To obtain the average contribution to the square Higgs field from these 
Yukawa fields around the Dirac sea top quarks - in the energy scale range 
around $\mu$ -  we shall then multiply 
this $(g_t/2)^2\frac{\sqrt{e} - 1/\sqrt{e}}{ 4\pi * \mu}$ with the density 
of Dirac sea top quarks $\frac{\mu^3}{2\pi^2}$ and by 6 from the number
 of internal states. So we end up with the product
\begin{equation}
``\phi_h^2\hbox{''}contribution|_{virtual\; tops} =(g_t/2)^2 
\frac{\sqrt{e} - 1/\sqrt{e}}
{4\pi *\mu} * \frac{\mu^3}{2\pi^2} * 6  =
(g_t/2)^2 \frac{6\mu^2(\sqrt{e}-1/\sqrt{e})}{(2\pi)^3}. 
\end{equation}
Now we have to note that the scale $\mu$ here should be essentially 
the top mass and thus we take it to be say 
\begin{equation}
\mu \approx g_t <\phi_h>.
\end{equation}
Inserting this we get our expression for the Dirac sea top quarks
in the scale range around $\mu$ 
to the Higgs field square average rewritten to
\begin{equation}
``\phi_h^2\hbox{''}contribution|_{virtual \; tops} = 
g_t^4 \frac{6(\sqrt{e}-1/\sqrt{e})}{32 \pi^3} * <\phi_h>^2\label{vtcontri}.
\end{equation}
%This expression and especially its behavior as $<\phi_h>$ is varied 
%has to be compared with the naive contribution of $<\phi_h>^2$ to
%the quantity $<\phi_h^2>$ which is what we are minimizing.
%Imposing that the change of the ``naive'' contribution under a scaling 
%with a factor $e$ should be equal to the just estimated virtual top 
%contribution we get a balance requirement equation:
%\begin{equation}
%g_t^4 \frac{6(\sqrt{e} - 1/\sqrt{e})}{32\pi^3}  = e-1/e
%\end{equation} 
%Crudely this implies that $g_t^4 $ be 400, or $g_t \approx 4 to 5.

%This expression is to be added to the ``trivial '' contribution to 
%give us the full ...I must think!

This were certainly very crude and one could easily imagine for instance 
an enhancement effect of the type that the top-quarks might attrackt 
each other and thus make their Yukawa fields strengthen each other so as to 
make the fluctuations in the Higgs field - meaning the contribution to 
the squared Higgs field - become bigger. Such an effect could make the value
of the Yukawa coupling needed at the weak scale for the balancing weaker.
According to Colin Froggatt and myself \cite{boundfirst}\cite{bound}
\cite{hierarchybound} there could even be 
a phase transition for some value of the top-Yukawa-coupling $g_t$ so
that a new phase of the vacuum, one with a condensate of a bound state 
of six top plus six antitop quarks,  would set in once the coupling $g_t$ 
at the  weak scale were below a certain limit which we estimated to be 
a critical value $g_{t \; critical}= 1.02\pm 14 \%$. If there is indeed such a
phase transition then very likely the minimum in the average of the 
squared Higgs field $<|\phi_h|^2>$ as a function of the unsquared 
average $<\phi_h>$
would occur just at the phase transition value of the top-Yukawa coupling.
 
\subsection{Discussion using a bound state condensate}
In any case it  is from dimensional arguments relatively 
easy to see that  the minmum in the average of the squared Higgs field as a 
function of the Higgs field average has to occur for some value of the 
running top-Yukawa coupling {\em which is of order unity}, only deviating 
by various factors $\pi$ etc. so that we can take our model to be at 
first successfull just from the mere fact that the experimental top
Yukawa coupling at the weak scale is indeed as needed of order unity. 
In fact it is 
$g_t=0.935$. If indeed the experimental coupling by nearer calculation
should turn out to be  as Froggatt and I claim, but disbuted by the 
Stony Brok group \cite{Stonybrook},  the phase transition coupling 
%is the one realized in nature at the weak scale, 
it would be a real support for the 
idea of the present article. It would namely  as already just said 
%it would 
very likely be the minimum in the squared Higgs field average just at the 
phase transition value. In that sense the minmization of the squared Higgs 
field 
average could automaticly directly lead to the existence of two degenerate 
phases in this special case of the phases with and without the 
6 top plus 6 antitop bound state condensate. But this would not be so new 
in the sense that Ninomiya and I already published an article 
\cite{SIMPP} suggesting that in general the model minimizing the 
imaginary part of the action $S_I$ gets allowed to even tune in some 
coupling constants would almost unavoidably lead to the ``multiple 
point principle'' \cite{MPP}. So the case of the degenracy of the bound state 
condensate vacuum and the one without - in which we presumably live - would
only be a special case of that. The present article argument for 
solving in a way the hierarchy problem by giving a mechanism leading to the 
finetuning of the Higgs mass scale to be exponentially small is also then 
not quite new in as far as we in our earlier work\cite{hierarchybound} used 
the multiple point 
principle as a supposed principle to argue for the top-Yukawa coupling to have 
to have the critical value at the weak scale, and used that to fix this weak 
scale. Rather would should consider it that I in the present article bring
 the 
same numerical coincidense - namely that the Yukawa coupling for the top-quark
at the weak scale is just the critical one causing the phase transition - 
in to confirm the complex action model in a slightly different and one could 
say more direct way. 

\subsection{Discussion using the crude calculation}

Although it is very likely that some phase transitionn could be involved in 
making there be a minimum in the average of the squared Higgs field 
$<|\phi_h(x)|^2>$ for $<\phi_h(x)> \ne 0$ we could also just continue 
our estimate using (\ref{vtcontri}). We would have to compare (\ref{vtcontri}) 
with the corresponding contribution to the average squared field 
$<|\phi_h(x)|^2>$ from the average simply comming in via $<\phi_h(x)>^2$.
Let us take from the range considered this contribution as 
\be
\phi_h^2 contribution|_{naive \ classical} = <\phi_h(x)>^2 (e - 1/e).
\ee 
Then the balance condition which should be satisfied in order for the 
field average $<\phi_h(x)>$ to be at the minimum we discuss would be

\begin{eqnarray}
``\phi_h^2\hbox{''}contribution|_{virtual \; tops}& = &
g_t^4 \frac{6(\sqrt{e}-1/\sqrt{e})}{32 \pi^3} * <\phi_h(x)>^2\\
 & =& <\phi_h(x)>^2 
(e-1/e)
 =\phi_h^2 contribution|_{naive \ classical}.
\label{balance}.
\end{eqnarray}

 This condition now is easily transformed into the condition
\be
g_t^4 *6= 32 \pi^3 (\sqrt{e} + 1/\sqrt{e}) 
\ee
From here we then get 
\be
g_t = ( 165.4 ( 1.649 +0.6065))^{1/4} = 373.0 ^{1/4}  = 4.39
\ee
This our predicted $g_t$-value is a bit bigger than the experimental
value $g_t|_{exp} = .935$, but taking the crudeness into account 
it is not so bad. If indeed as suspected from the bound state speculations 
there were an effect of the virtual top-quarks clustering together the 
Higgs field surrounding a cluster of virtual tops 
would gets its {\em square } of its strength - as we used above - increased 
by a factor being the effective number of virtual tops in the cluster. 
This factor would in our estimate come to decrease our value for the fourth 
power of the $g_t$ by the number of virtual tops in the cluster.
If we for example optimistically could have the effective number 
of tops in the cluster being the 12 that were the number of constituents 
in the bound states of 6 top + 6 anti tops we would decrease our predition for 
$g_t$ by the fourth root of this 12. That would bring the prediction down 
from our first 4.39 to 2.35. It is still not perfect but really we cannot 
expect bettter accuracy with the just made estimate. 
% surrounding Higgs field 
 
In any case it is now surely a possibility that it were somehow arranged that 
the imaginary part of the action were minimized as far as the vacuum 
Higgs field square is concerned. This clearly supports our complex action 
model.
  
\section{Worrying about too much arrangement of the state of the universe}
\label{toomuchque}

%This is to be inserted into the file complexsupport3.tex :

There is a severe problem with our complex action model: If really 
the integral $\int |\phi_h|^2 d^4x $ of the Higgs field square 
integrated over all space time should be so important in fixing 
the history of the universe that even a moderate number of Higgses 
that would have been produced in the SSC\cite{SSC} machine should have 
effectively 
influenced  the political history of this machine, then the much 
more common particles that are surrounded by Higgs fields like the 
quarks and the leptons would make much more dramatic contributions
a priori. It thus seems at first that depending on how the sign of the 
relative extra Higgs square contrbutions caused by such more common 
reactions transforming particles,
%particles reactions 
that can change the Higgs square field around them, the total set of 
particles involved would be pushed completely to one side in order to 
minimize the integrated Higgs field squared. Let us to be concrete 
look at neutron decay. It is the reaction
\begin{equation}
n <--> p + e + \bar{\nu}_e.\label{process}
\end{equation}   
It is now not really so difficult to calculate at least 
crudely the difference in the total change in the Higgs field square 
integrated over space time - or better just over space - when 
the above process takes place. Depending on the sign of the 
shift in the Higgs field square caused by the process(\ref{process}) it should 
after 
our complex action theory be arranged if at all possible - by adjustung 
the initial condition and the coupling constants allowing adjustment -
to be so that the process were driven (almost) totally to one side.
If for instance 
 it turned out that the neutron by its interaction with the Higgs field 
made the Higgs field squared diminish more around the neutron than the 
corresponding deminishing around all of its three decay products added up,
then there should be no decay product combinations left and somehow 
all should have been arranged now to be in the form of neutrons.
This would be very difficult indeed to arrange by appropriate initial 
conditions, since how could even a ``God'' able to arrange initial conditions 
keep neutrons stabilized against decay if there were no protons left 
to keep them bound into nuclei? If on the other hand the difference 
should have been 
so that the decay products of the neutron would be more favorable 
by suppressing the Higgs field square than the neutron, then there should 
have been arranged to be no neutrons left. That would mean it should all 
be hydrogen, also not so easy to keep, for would there not develop stars?
And the stars would presumably make helium etc. with lots of neutrons inside.

None of the two scenarios are very easy to realize, and certainly none 
of them are true experimentally! At first it seems that we here have a severe 
problem for the complex action theory! There is, however, one way out:
If it had been arranged that the coupling constants were just tuned in to 
be so that the amount of suppression of the Higgs field square were just 
the same by the set of all the three decay products and by the neutron itself,
then there would be no need for bringing the process to one side or the other.
.It were in fact the main point of my paper connected with my talk in Spaatind 
\cite{Spaatind} that one could expect that our complex action theory could 
easily be expected to precisely arrange such an equallity of the Higgs 
field square suppression from the particles on the two sides of
a transition  equation, such as (\ref{process}). 
%as the one for the neutron decay. 
Indeed I claimed    
to find approximate {\em experimental evidence} for the relations between 
quark masses 
proton mass etc. being deduced from requiring this relation to be 
true. That is to say I claimed that it had somewhat miraculously, one 
could almost say, been arranged that whether the neutron decays or not, 
will make no change in the Higgs fields square integral over space. By this 
almost miraculous adjustment the need for having the reaction pushed 
almost totally to one side in our model gets dispenced of. This balance 
making the neutron decay irrelevant for the Higgs field square integral 
is what rescues our model for the problem with this reaction, neutron decay.
But there are other reactions that could threaden our model by having 
to be pushed quite to one side as far as possible or by having associated 
with them another fine tuning of couplings arranging, that the reaction in 
question just does not cause change in the Higgs field square suppression.
A number of relations of this type might be found, and it would of course 
be very interesting, if they are fulfilled  in nature.
 
\section{The relation}
\label{relation}
So far I only looked properly on the relation expected to avoid that the 
neutron decay reaction should  be pushed completely to one side.
To write that needed relation I did estimate in the article \cite{Spaatind}
the amount of change in the Higgs field square around the neutron and on the 
other hand the sum of the changes around the three decay products, the proton,
the electron and the neutrino. The relation we should derive is then the 
relation needed to  keep the integrated square of the Higgs field the same 
before and after the neutron decay, so that letting the neutrons decay does 
not matter for the value of $S_I$ which is the space time integral of the 
Higgs field  square. 

\subsection{For the Relation Calculational needed Assumptions}

Let me here make a few remarks about this calculation:

\begin{itemize}
\item{1)} We know that the Higgs Yukawa field around a quark or a lepton 
is proportional  to the Yukawa couling for that quark or
 lepton, or W etc.

\item{2)} and we suppose that we can Taylor expand so as to argue for that 
the change in this {\rm square} of the Higgs field is indeed also proportional 
to the Yukawa coupling. This is actually not so obvious for two reasons,
but we still shall use this approximation:
\begin{itemize}
\item{A)} We strictly speaking use that the change in the Higgs field 
around the 
quark say is small compared to the Higgs VEV(=the vacuum expectation 
value). This is o.k. for the light quarks and the electron in the dominant 
region in their Yukawa potentials.

\item{B)} The idea also mentioned in this article that the Higgs vacuum 
expectation 
value should precisely be the one leading to a minimum in the Higgs field 
square, stricly speaking tells that precisely that derivative (of 
the square w.r.t. the field itsef),  we assume to dominate the Taylor 
expansion 
in calculating the effect of the Yukawa field around a quark say, is zero.
%\begin{itemize}
There may though be some excuses for not taken that too seriously:
\begin{itemize}

\item{B1)} The Yukawa potential is not constant in space so that spatial 
Fourier componnts relevant are not exactly that $\vec{p}=0$ component, 
for which the derivative is zero.

\item{B2)} As we speculate there could likely be an, although weak, first 
order phase transition just at the minimum for $<|\phi_h|^2>$ so that 
there is at least slightly non-zero derivatives from right and from left,
but no true zero-slope place. Then the sign  of the derivative would 
depend on from which of the two meeting phases we work. The speculation is 
that the phase relevant for the Higgs field average below the minimum point 
is a phase with a condensate of bound states, while we above have the 
``normal'' phase without such a condensate. It should then be the right 
derivative, meaning the one in the ``normal'' phase that we shall use in the 
Taylor expansion.   
  \item{B3)} One could also expect an indirect effect of the Higgs field 
square comming as a loop correction via an effective complex kinematical 
term coefficieint.
\end{itemize}

\item{C)} So from A) and B) the change in the Higgs field square integrated 
over 3-space is proportional to the Yukawa coupling of the particle 
that makes the change.

\item{D)} The integral over space of the Higgs field (squared or not)
is in addition proportional to the inverse of the $\gamma$ for the particle.
We could say it simply reflects the Lorentz contraction, correcting the 
volume crudely covered by the Yukawa potential by the factor $\gamma^{-1}$.

\item{E)} The appearance of the just mentioned Lorentz contraction factor 
means that the neutrino, which in practice allways runs with almost 
exactly the 
speed of light (even if there are extremely tiny neutrino masses different 
from  0.)gets totally Lorentz contracted, 
and so the neutrino can be left out, so that we only have to consider it as if 
the neutron decayed into an electron and a proton alone.
%(ignoring the neutrino)

\item{F)} Thinking as is typically what happens that we have to do with 
non-relativistic neutrons, protons and elctrons in the bulk of the universe,
we have no Lorentz contraction factor for the elctron, but even though the 
proton and the neutron are essentially at rest the quarks inside them move 
with speeds comparable to that of light.

\item{G)} It is indeed the major calculational trouble of making our relation 
to estimate the appropriate average of the $\gamma^{-1}$ factors for the 
quarks inside the proton and the neutron. 

\begin{itemize}
\item{G1)} In crudest approximation for the average of the correction factor for 
Lorentz contraction of the  Yukawa-field around a quark is $<\gamma^{-1}> 
\approx <\gamma>^{-1} = <E_q /m_q >^{-1} $ where $E_q$ is the energy 
of the quark inside the, say, proton and $m_q$ its (current algebra) mass.

\item{G2)} In the articel \cite{Spaatind} I gave a name 
\begin{equation}
``ln'' = <\gamma><\gamma^{-1}> \label{lndef}
\end{equation}
  to the correction to this just mentioned approximation, and in this article 
are mentioned estimates like 
$``ln'' = \frac{\ln{2<\gamma>}}{2- <\gamma>^{-1}}$,
 which for 
$<\gamma> = 55$ would give $``ln'' =  2.37$ while for $<\gamma> = 27.5$ it gives 
$``ln'' = 2.05$.

\item{G3)} Another difficulty that needs further estimation is to estimate 
how big a fraction of say the  proton mass is actually sitting as energy 
of the 
valence quarks, so that we can use one third of the amount as the average 
$<E_q> $ of the quark energy $E_q$ in the proton rest frame. It is presumably 
not quite unreasonable to estimate that about {\em half the proton mass} sits 
as 
energy on the three valence quarks together, so that we can take (crudely of 
course) $<E_q> \approx m_{proton}/(2*3) \approx 160 MeV$.  

\item{G4)}(Current algebra) quark masses are of course not extremely well 
determined for the light quarks $u$ and $d$, with which  we are here concerned,
 but we may take:
\begin{eqnarray}
m_u & = & 1.7  \hbox{to} 3.3 MeV\\
m_d & = & 4.1 \hbox{to} 5.8 MeV\label{qmasses}
\end{eqnarray}
  
\end{itemize}
\end{itemize}
\end{itemize}

\subsection{Agreement of our relation involving quark masses etc} 
From this proceedure I then arrived in article \cite{Spaatind} to the 
relation 
\begin{equation}
\sqrt{m_d^2 - m_u^2} =\sqrt{E_q m_e}/``ln''\label{relationf}
\end{equation}
which is  relativly well satisfied, if we take the quark masses 
(\ref{qmasses}), $E_q \approx 160 \ MeV$ and $``ln'' =2.3_7 $.
In fact then we would get (using $m_e = 0.511\  MeV$) 
%\end{document}
\begin{eqnarray}
 R.H.S.& = &\sqrt{E_qm_e}/``ln'' = 3.81 \ MeV \\
L.H.S.& =& \sqrt{m_d^2 - m_u^2} = \sqrt{13._9} \hbox{to} \sqrt{22._75} \ MeV\\
& = &
3.7_3 \hbox{to} 4.7_7 \ MeV. 
\end{eqnarray}

Really it should be possible to extract a better estimate for the 
$<\gamma^{-1}>$ and $<\gamma> $ from using the parton distribution 
function(PDF) 
for the nucleons. Actually looking at the PDF's for the light valence 
quarks one sees a peak at the Bjorken\cite{Bj} momentum fraction $x=0.2_0$
\cite{PDF}. That 
should be taken to mean that 
the typical energy of a valence quark, $u$ or $d$,  should be 
$E_q = 0.2_0 m_N = 0.19 \ GeV 
= 190 \ MeV $. Inserting this $E_q$-value instead of the $160$ Mev 
would bring the right hand side of our relation up to
\begin{equation}
R.H.S. = \sqrt{E_q m_e}/``ln'' = 4.15 \ MeV.
\end{equation}
This is very much inside the range of uncertaity of the quark masses, so
our agreement is very good! 

If the (current algebra) quark masses were known better it would be 
worthwhile to evaluate 
say $<\gamma^{-1}>$ more accurately from using PDF's and/or other informations
so as to test our relation more accurately.

\section{A Few Further Evidences for the Minimization of the Integral of the 
Higgs Field Squared}
\label{further}

\subsection{Why Nuclear Binding Ends up Small Compared to the Kinetic and 
Potential Energies Seperately}
\label{binding}

Similarly to the above discussed worry, that in our model there should either 
have been no neutrons or no possibilities for combining a proton, an electron, 
and an antielectron neutrino, we shall also ask, if there ought to have been 
either only nucleons bound into  heavier nuclei or only unbound nuclei 
meaning only hydrogen. The point is that, if the suppression of the Higgs 
field square 
by a bound and an unbound nucleon is not the same, then the principle 
of minimizing the square of the Higgs field integrated over space and time 
would either lead to an arrangement, so that there were only bound nucleons 
or to one, in which there were only unbound nucleons. Providing our already 
derived relaton (\ref{relation}) it should not matter for this purpose whether 
the nucleon is a proton or a neutron (corrected for the accompagying electron
for charge neutrality the proton (with its electron) and the neutron would 
namely suppress the Higgs field equally much). 

Our complex action model would now have a problem unless it happens by 
some finetuning or for whatever - ``arranged'' - reason that the Higgs field 
square decrease due to the precense of a  nucleus is the same integrated over 
space as if we have the same nucleons seperated out as free nucleons. Thus we 
must 
now at least very crudely estimate how it should be in order that these two
suppression amounts should indeed become the same number for 
$ \int |\phi_h|^2 d^3\vec{x}$, i.e. this quantity would be changed equally 
much by the nucleus as by its constituents being free (and with low, 
nonrelativistic, speeds).

To make a first very crude estimate of the difference between the two 
suppression amounts we remark, that assuming the quantity (\ref{lndef})
not to vary significantly because of the binding taking place the fulfilment
of the requirement amounts to that the energy carried by the quarks 
(kinetic and mass energy of the quarks) should be the same whether the nucleons
are bound or not.       
 In the approximation that the energy of the non-quark energy nature 
(meaning say the gluon carried energy)  being constant or proportional
to that for the quarks we would then say that in the very crudest 
approximation the energy change by the baryons being bound into heavier nuclei 
should zero, i.e. small. The degree of smallness is of course given by the
deviation from correctness of the assumtions we made here. Let us take the 
smallness to mean though, that the binding energy that should be small should
be that  compared to the kinetic energy or the potential energy seperately. 
Interpreted this way that is in fact what somewhat surprisingly is wellknown
to be the case for nuclear forces. In spite of the Fermi kinetic energy of the 
the nucleons in the nuclei and the potential energy being  both 
rather large, the Fermi kinetic energy being 38 MeV per nucleon
\cite{Fermienergy} 
typically  the NET binding energy per nucleon 
is ONLY in the most stringly bound region of nucleides 8.8 MeV\cite{Fewell}. 
This means the net binding is down by a factor $38/8.8 =4.3$.
That is to 
say that in the first 
approximation  there is indeed very crudely  an approximate cancellation 
so that if as we assumed crudely the energy of the quarks follow that 
of the nucleons (proportionally) in such a way that even the potential 
energy of the nucleons gives rise also to kinetic and mass energy of the 
quarks, then the quark energy would not change by the binding. But that in 
turn were what we needed to not get contradictionally to truth
that either all nucleons should be bound or oppositely none being bound.

It would be extremely interesting for checking our model 
to estimate more accurately how the gluonic part of the energy gets 
changed by binding the nucleons into nuclei and how our crude estimate should 
in addition be corrected. Then one could namely hope to calculate what precise 
value of the binding energy (compared to say the kinetic energy) would be
required for our model to be consistent. It would of course be an interesting 
victory, if such a calculation would lead to the experimental binding energy
ofthe nuclei.

\subsection{Could we even Predict the Outcome of the Debate on Nuclear Power?}
One would expect that the nuclear power production processes a priori 
would change the bindings of the nuclei in detail. So although the binding 
of nuclei to very firt approximation may well have been tuned in to 
make no change in the integrated Higgs field squared under such processes,
it would presumably require an extra little fie tuning to get the Higgs field 
square integrated over space totally independent of whther nuclear power 
is being used or not. If we can calculate the difference in change of the 
Higgs field square integrated over space for the nuclear power fuel 
relative to the nuclear power waste, we should be able to 
estimate how the squared Higgs field would be arranged to be the smallest 
integrated over space time, by having nuclear power being used or not.
If we for example find that the waste had the smallest Higgs field square 
integral, then our model would predict that by ``almost miraculous accidents'' 
the political development would be so as to make nuclear power come to 
be used. If we get the opposite result that it is fuel that gives the lowest 
integral of the square of the Higgs field, then the use of nuclear power 
should be ``almost miraculously by accident'' stopped.

\subsection{Higgses Produced in Cosmic Ray Impacts}

In the earlier article \cite{Spaatind} I argued for that using 
e.g. theoretical estimates of Tully \cite{Tully} and extrapolating them 
linearly 
as the cross section being a linear  function of the center of mass 
energy $\sqrt{s}$ of the colliding nucleons one 
could define crudely ``an effective Higgs production threshold'' 
as that value for the center of mass energy $\sqrt{s}$ for which 
the extrapolated cross section passes zero. Extrapolating this 
``effective threshold for Higgs production'' as a function of 
%would get for a 
the Higgs mass 
to the  range favoured by our own prediction \cite{Higgsprediction} of the 
Higgs mass $ M_h = 120 \ GeV$ an effective threshold of $\sqrt{s}|_{threshold}
 =2.7 \ TeV $ is reached. That in turn when translated to scattering on 
a fixed target
as a proton comming in and hitting the atmosphere gives a Higgs production 
effective threshold around $3.6 * 10^{15} \ eV$ and this is very close 
indeed to the ``knee'' in the Gaisser curve of the cosmic ray intensity 
as a function of the energy at about $2 * 10^{15} \ eV$. That means that 
the fall off with extra speed of the cosmic ray intensity signaled by this
``knee'' is as if it were just ``chosen'' to avoid producing Higgses 
effectively! It is even so that especially just above the ``knee''
the cosmic radiation seems more rich in presumably iron ions than the 
lighter elements supposedly dominating below the ``knee''.For iron of course 
the ``effective Higgs threshold'' would be higher in energy by a factor 
of the order of the ratio of the iron to light nuclei atomic weight. 
at extremely high cosmic ray energies there is even the ``angel''
(the spectrum flattens off again) so there 
are some although remarkably few cosmic rays producing Higgses. Thus our 
model does NOT fully prevent Higgs production in cosmic ray, if it were true.  

The coincidence order of magnitudewise of the ``knee'' with the ``effective
Higgs production threshold'' (for an almost by now unavoidable Higgs mass 
range) would be yet an interesting coincidence supporting  the principle of 
the 
Higgs field squared  being kept minimal, since of course the precense of 
genuine 
Higgs particles means an increase in the Higgs field square. 

Really of course it would have beenbetter in supporting the complex action 
model to have no cosmic ray above the knee at all, but we only got the ``knee''
and then there even were the ``angel'', which however very likely consists of 
cosmic rays from further away galaxes, while the cosmis ray below the ``knee''
is likely from our own galaxy. So the fraction of the very high energy cosmic 
ray particles actually hitting any atmosphere or any astronomical object 
would be lower than for the particles below the ``knee'.

\section{Conclusion of much evidence for the complex action indirectly!}
\label{conclusion}

The new result of the present article is the claim that it is possible 
that the surprizingly little but not - even comapared to Planck scale - 
totally zero expectation value of the Higgs field $ <\phi_h> = 246 \ GeV$
is indeed that value of this expectation value which leads to the minimal value
of the related average $<\phi_h^2>$ of the {\em square} of this same 
Higgs field. We imagine indeed that there is some parameter of the vacuum or 
say some coupling constant that can at least be imagined to be 
 variable somehow and which thereby leads to a correlated variation of 
both the average and the square of the average of the Higgs field. 
I.e. we shall think of both $<\phi_h> $ and $<\phi_h^2>$ as depending on 
the same such parameter. Then our claim that basically inside the 
Standard Model
- except for this ```variability'' -   
the parameter has that value which makes the average of the {\em square}
the smallest possible! Indeed we claimed that this minimal average of the 
square is achieved - since its for us most interesting dependence comes
from fluctuations due to virtually present top-quarks - when the running 
Yukawa coupling for the top-quark has a certain specific value, which is of 
order unity, but which we could not so far calculate so accurately as 
would have been usefull. It is nevertheless if one accepts some physical 
model/story behind a solution to the problem of the large scale ratio of the 
weak scale to say the Planck scale. That is a solution very similar to the 
one which C. D. Froggatt, L. V. Laperashvili and earlier presented 
\cite{hierarchybound}. The crux of the matter is that by relating the 
weak scale 
to a {\em running} coupling - in the present work the top-Yukawa coupling - 
it gets naturally of an exponential size, because the running of 
couplings are ``logarithmic'', logarithmically slow.

The story behind the principle of minimizing the average of the 
square of the Higgs field, which we used, is the ``complex action
model'' of Ninomiya and myself. In this model we namely arrive to 
minmizing the imaginary part of the action $S_I[history]$ calculated 
for the actual history $history$ of the Universe. We then make the 
approximation that the vacuum dominates and that the mass square 
term in the imaginary part of the action $S_I$ dominates. Thus the
average square of the Higgs field multiplied by the space time volume 
becomes approximately the $S_I$ and thus minized when the average of squared 
field is minimal. Thus the average of the (unsquared) Higgs field should
get adjusted to minimize the average of the squared field.

We reviewed an earlier work of mine also made on the hypotesis that 
the average of the square of the Higgs field should be minimized:

If say the neutron versus the decay product of a neutron 
made different change in this squared Higgs field integrated over all space 
and time, then the most bennificial in the sense of the smallest 
Higgs field squared integrated, would either be a history with no 
neutrons at all, or one with no decay products of neutrons meaning no 
elctrons and protons. But in Nature we do find that the reaction of 
neutrondecay is in this way pushed to one side or the other. Rather there
are Nature, both combinations of electron and protons and even antineutrinos,
{\em and} neutrons! The reaction is {\em not} pushed to one side.
According to our minimization of $S_I$ principle this reaction not being 
pushed to a side is {\em only possible provided the modification of the 
Higgs field (square) by a neutron and by its decay products is the same} !

Then I rewied a formula needed for this modification of the Higgs field 
squared and integrated be the same for the neutron and for its decay products,
and love and behold this relation is fullfilled inside our estimation 
accuracy! 

We should consider these two numerical agreements of the priniple of 
minimizing the integral  over space and time of the Higgs field square 
as a success for the hypotessi that such a quantity were indeed 
{\em minimized} ! 

The two cases thus argue for that some parameters of nature are indeed 
adjusted towards a minimization of some quantity like 
the intergal over space time of 
the Higgs field squared 
and integrasted.  
   
If we include the more weak agreements we have further somewhat supporting 
successes mentioned above:

1) The smallness of the binding of nuclei compared to say the kinetic energy 
of the nucleons inside the nuclei could be adjusted so as to make the change 
in the over space integrated Higgs field (squared) be independent to 
whether the nucleons are bound or not. We can consider the fact that the 
binding energy of nuclei is relatively small compared to the Fermi-energy 
and the potential energies seperately as suggestively making it more likely 
that the suppression of the Higgs field squared could have been finetuned to 
be the same whether the nucleon is bound or not.

2) The ``knee'' in the cosmic ray intensity curve versus energy 
is ``made'' to cut away Higgs production in the atmosphere of astronomical 
objects being hit by the cosmic radiation.

3) The bad luck of the S.S.C. machine\cite{SSC} that would have produced 
a lot of 
Higgses, but which were killed by the Congress of the United States.

If we further take it that an imaginary action theory could be used to 
argue for the Multiple Point Principle (MPP)\cite{MPP}, then  successes of this
principle, that there should be many degenerate vacua - or rather one vacuum 
should be just about to decay into another one (called 
Meta-MPP\cite{MetaMPP}) -,
could be counted as successes of the complex action model too.
Now the  claims of success of such a principle of essentially 
degenerate vacua counts a Higgs mass as low as allowed in the Standard 
Model, a 
prediction seemingly getting slowly more and more likely as higher masses 
for the Higgs get excluded. We have also claimed that it could give the 
top-Yukawa-coupling if we believe in the possibility of forming certain
bound states of six top and six antitop quarks. We even claimed already that 
 this MPP could ``solve the problem of why the weak scale is exponentially 
small compared to the say Planck scale''. The main point of the present 
article is really a slightly different version of the same prediction
- an exponentially small weak scale - but in a somewhat different setting 
in details, although both ways of solving this scale problem comes directly 
(this article) or indirectly (our earlier articles with Laperashvili and 
Froggatt\cite{hierarchybound}) from the complex action model.

Taking just some of these ``evidences'' for the complex action 
\cite{ownmMPP}\cite{SIMPP} 
seriously 
would mean that there {\em is} some remarkable evidence for the truth 
of such a model or some similar model with similar predictions, especially 
a minimization of the Higgs field squared would be called for.


\begin{thebibliography}{99}
\bibitem{ownmMPP}
%\cite{Nielsen:2008zz}
%\bibitem{Nielsen:2008zz}
  H.~B.~Nielsen and M.~Ninomiya,``Nonexistence of irreversible processes 
in compact space-time,''
  Int.\ J.\ Mod.\ Phys.\  A {\bf 22} (2008) 6227.
  %%CITATION = IMPAE,A22,6227;%%
%\cite{Nielsen:2008cm}
%\bibitem{Nielsen:2008cm}
  H.~B.~Nielsen and M.~Ninomiya,
  ``Test of Influence from Future in Large Hadron Collider: A Proposal,''
  arXiv:0802.2991 [physics.gen-ph].
  %%CITATION = ARXIV:0802.2991;%%
%\cite{Nielsen:2007mj}
%\bibitem{Nielsen:2007mj}
  H.~B.~Nielsen and M.~Ninomiya,
  ``Complex Action, Prearrangement for Future and Higgs Broadening,''
  arXiv:0711.3080 [hep-ph].
  %%CITATION = ARXIV:0711.3080;%%

%\cite{Nielsen:2007ak}
%\bibitem{Nielsen:2007ak}
  H.~B.~Nielsen and M.~Ninomiya,
  ``Search for Future Influence from L.H.C,''
  Int.\ J.\ Mod.\ Phys.\  A {\bf 23} (2008) 919
  [arXiv:0707.1919 [hep-ph]].
  %%CITATION = IMPAE,A23,919;%%


%%\cite{Nielsen:2007kz}
%%\bibitem{Nielsen:2007kz}
%  H.~B.~Nielsen and M.~Ninomiya,
%  ``Degenerate vacua from unification of second law of thermodynamics with
%  %other laws,''
%  arXiv:hep-th/0701018.
%  %%CITATION = HEP-TH/0701018;%%

%\cite{Nielsen:2006pz}
%\bibitem{Nielsen:2006pz}
  H.~B.~Nielsen and M.~Ninomiya,
  ``Future dependent initial conditions from imaginary part in lagrangian,''
  arXiv:hep-ph/0612032.
  %%CITATION = HEP-PH/0612032;%%

%\cite{Nielsen:2006td}
%\bibitem{Nielsen:2006td}
  H.~B.~Nielsen and M.~Ninomiya,
  ``Trouble with irreversible processes in non-boundary postulate. and  
perfect
  match of equation of motions and number of fields,''
  arXiv:hep-th/0602186.
  %%CITATION = HEP-TH/0602186;%%

%\cite{Nielsen:2006th}
%\bibitem{Nielsen:2006th}
  H.~B.~Nielsen and M.~Ninomiya,
  %``Compactified time and likely entropy: World inside time machine: Closed
  %time-like curve,''
  arXiv:hep-th/0601048.
  %%CITATION = HEP-TH/0601048;%%

%\cite{Nielsen:2006vc}
%\bibitem{Nielsen:2006vc}
  H.~B.~Nielsen and M.~Ninomiya,
  %``Intrinsic periodicity of time and non-maximal entropy of universe,''
  Int.\ J.\ Mod.\ Phys.\  A {\bf 21} (2006) 5151
  [arXiv:hep-th/0601021].
  %%CITATION = IMPAE,A21,5151;%%

%\cite{Nielsen:2005ub}
%\bibitem{Nielsen:2005ub}
  H.~B.~Nielsen and M.~Ninomiya,
  ``Unification of Cosmology and Second Law of Thermodynamics: Solving
  Cosmological Constant Problem, and Inflation,''
  Prog.\ Theor.\ Phys.\  {\bf 116} (2007) 851
  [arXiv:hep-th/0509205].
  %%CITATION = PTPKA,116,851;%%

\bibitem{SIMPP}
%\cite{Nielsen:2007kz}
%\bibitem{Nielsen:2007kz}
  H.~B.~Nielsen and M.~Ninomiya,
  ``Degenerate vacua from unification of second law of thermodynamics with
  %other laws,''
  arXiv:hep-th/0701018.
  %%CITATION = HEP-TH/0701018;%%

\bibitem{FH} Feynman and Hibbs,

Quantum Mechanics and Path Integrals (Hardcover)
by Richard P. Feynman (Author), A. R. Hibbs (Author)



\bibitem{Wentzel} Wentzel,

Physics Letters A
Volume 324, Issues 2-3, 12 April 2004, Pages 132-138; 

Salvatore Antoci and Dierck-E. Liebscher,
 Wentzel's Path Integrals,International Journal of Theoretical Physics
Publisher       Springer Netherlands
ISSN    0020-7748 (Print) 1572-9575 (Online)
Issue   Volume 37, Number 1 / January, 1998
DOI     10.1023/A:1026628515300
Pages   531-535
Subject Collection      Physics and Astronomy
SpringerLink Date       Wednesday, December 29, 2004;



International Journal of Theoretical Physics, 
Volume 37, Number 1, 1 January 1998 , pp. 531-535(5)

\bibitem{Dirac}
  Dirac, The principles of Quantum Mechanics, 1930 


\bibitem{old}
%\bibitem{nonlocalfirst}
%\cite{Bennett:1995ag}
%\bibitem{Bennett:1995ag}
  D.~L.~Bennett, C.~D.~Froggatt and H.~B.~Nielsen,
  ``Nonlocality as an explanation for fine tuning in nature,''
  (CITATION = C94-08-30);

%\bibitem{nonlocalsec}
%\cite{Bennett:1994yx}
%\bibitem{Bennett:1994yx}
  D.~L.~Bennett, C.~D.~Froggatt and H.~B.~Nielsen,
  ``Nonlocality as an explanation for fine tuning and field replication in
  nature,''
  arXiv:hep-ph/9504294.
  (CITATION = HEP-PH/9504294;)


\bibitem{vacuumbomb}
%VACUUM BOMB 
%\bibitem{vacuumbomb1sv}
 D.~L.~Bennett, ``Who is Afraid of the Past'' ( A resume of discussions 
with H.B. Nielsen during the summer 1995 on Multiple Point Criticallity 
and the avoidance of Paradoxes in the Presence of Non-Locality in 
Physical Theories), talk given by D. L. Bennett at the meeting of the
Cross-displiary Initiative at Niels Bohr Institute on September 8, 1995.
QLRC-95-2.
%\bibitem{vacuumbomb2th}
%\cite{Bennett:1996hx}
%\bibitem{Bennett:1996hx}
  D.~L.~Bennett,
  ``Multiple point criticality, nonlocality and fine tuning in fundamental
  physics: Predictions for gauge coupling constants gives alpha**(-1) =  136.8
  +- 9,''
  arXiv:hep-ph/9607341.
  (CITATION = HEP-PH/9607341;)
%\bibitem{vacuumbomb3co}
%\cite{Nielsen:1995rs}
%\bibitem{Nielsen:1995rs}
  H.~B.~Nielsen and C.~Froggatt,
  ``Influence from the future,''
  arXiv:hep-ph/9607375.
  (CITATION = HEP-PH/9607375;)

%\bibitem{landscape}
%%\cite{Carroll:2005ah}
%%\bibitem{Carroll:2005ah}
%  S.~M.~Carroll,
%  %``Is our universe natural?,''
%  arXiv:hep-th/0512148.
%  %%CITATION = HEP-TH/0512148;%%

%L. Smolin, "Did the universe evolve?," 
%Classical and Quantum Gravity {\bf 9}, 173- 191
%%Gâ€"
%%@191 
%(1992). 
%L. Smolin, The Life of the Cosmos (Oxford, 1997)
%
%Susskind, The cosmic landscape: string theory and the illusion of 
%intelligent design (Little, Brown, 2005). 
%M. J. Rees, Just six numbers: the deep forces 
%that shape the universe (Basic Books, 2001). R. Bousso and J. Polchinski, 
%"The string theory landscape", {\bf Sci. Am. 291}, 60
%%%Gâ€"%@
%- 69 (2004).
%%
%%
%
%S. Weinberg, "Anthropic bound on the cosmological constant", 
%Phys. Rev. Lett. 59, 2607 (1987).



%\bibitem{qfp}
% C. T. Hill, Phys. Rev. D24 (1981) 691;
%C. T. Hill, C. N. Leung and S. Rao, Nucl. Phys. B262 (1985) 517.
% B. Pendleton and G. G. Ross, Phys. Lett. 98B (1981) 291.
% W. Zimmermann, Commun. Math. Phys. 97 (1985) 211;
%J. Kubo, K. Sibold and W. Zimmermann, Phys. Lett. B200 (1989) 191	


%\bibitem{Fermienergy}
%Wikipedia, ``Fermi energy''.
%
%\bibitem{Fewell}
%Fewell, M. P. (1995). "The atomic nuclide with the highest mean binding 
%energy"American Journal of Physics {\bf 63} (7): 653 - 658. 
%Bibcode 1995AmJPh..63..653F 
%(http://adsabs.harvard.edu/abs/1995AmJPh..63..653F). doi:10.1119/1.17828 
%(http://dx.doi.org/10.1119\%2F1.17828)





%\bititem(Knee}
%See figure 2.2 of Sokolsky, P. 1989. Introduction to Ultrahigh 
%Energy Cosmic Ray Physics. Redwood City: Addision-Wesley Publishing Company

\bibitem{babyBanks}
Banks, T., 1988, Proleoma to a theory of bifurcating universes: a nonlocal 
solution to the cosmological constant problem, or a little lambda goes back 
to the future. Santa Cruz preprint SCIPP 88/09 

\bibitem{babyColeman}
Coleman, S. 1988a Why is there nothing rather than something : a theory of 
the cosmological constant. Havard prprint HUTP-88/A022.
Coleman,S. 1988b Black holes as red herings: topological fluctuations and 
the loss of quantum coherence. Nucl. Phys.B {\bf 307},864.

\bibitem{babyHawkings}
Hawking,S.W. 1984, The cosmological constant is probably zero,Phys.Lett. {\bf
B 134},403; 1987 Coherence down the wormhole, Phys. Lett.{\bf b 195},337; 
1988 Wormholes in space time 
Phys.Rev.{\bf D 37}, 904. 



\bibitem{HLbaby}
Hawking and La Flamme,
Physics Letters B
Volume 209, Issue 1, 28 July 1988, Pages 39-41
doi:10.1016/0370-2693(88)91825-4 | How to Cite or Link Using DOI
Copyright © 1988 Published by Elsevier Science B.V. All rights reserved.
  Permissions \& Reprints


\bibitem{selfbaby}

%\cite{Nielsen:1988cc}
%\bibitem{Nielsen:1988cc}
  H.~B.~Nielsen and M.~Ninomiya,
  ``BABY UNIVERSES, FINE TUNING PROBLEMS: A THEORY OF EVERYTHING ROBBING THE
  THRONE BY KILLING THE RIVALS,''
  %%CITATION = C88/10/17.2;%%

%\cite{Nielsen:1988kf}
%\bibitem{Nielsen:1988kf}
  H.~B.~Nielsen and M.~Ninomiya,
  ``A SOLUTION OF STRONG CP PROBLEM IN BABY UNIVERSE THEORY,''
  Phys.\ Rev.\ Lett.\  {\bf 62} (1989) 1429.
  %%CITATION = PRLTA,62,1429;%%



Nielsen, H. B.; Ninomiya, M.
Baby universe theory., by Nielsen, H. B.; Ninomiya, M.. Niels Bohr Inst., 
Copenhagen (Denmark), Dec 1989, 46 p., Paper presented at the 3. Regional 
Conference on Mathematical Physics, Islamabad (Pakistan), 18 - 24 Feb 1989.





%\cite{Nielsen:2010gq}
\bibitem{Spaatind}
  H.~B.~Nielsen,
  %``Remarkable Relation from Minimal Imaginary Action Model,''
  arXiv:1006.2455 [physics.gen-ph].
  %%CITATION = ARXIV:1006.2455;%%

\bibitem{landscape}
%\cite{Carroll:2005ah}
%\bibitem{Carroll:2005ah}
  S.~M.~Carroll,
  %``Is our universe natural?,''
  arXiv:hep-th/0512148.
  %%CITATION = HEP-TH/0512148;%%

L. Smolin, "Did the universe evolve?," 
Classical and Quantum Gravity {\bf 9}, 173- 191
%Gâ€"
%@191 
(1992). 
L. Smolin, The Life of the Cosmos (Oxford, 1997)

Susskind, The cosmic landscape: string theory and the illusion of 
intelligent design (Little, Brown, 2005). 
M. J. Rees, Just six numbers: the deep forces 
that shape the universe (Basic Books, 2001). R. Bousso and J. Polchinski, 
"The string theory landscape", {\bf Sci. Am. 291}, 60
%%Gâ€"%@
- 69 (2004).


S. Weinberg, "Anthropic bound on the cosmological constant", 
Phys. Rev. Lett. 59, 2607 (1987).



\bibitem{qfp}
 C. T. Hill, Phys. Rev. D24 (1981) 691;
C. T. Hill, C. N. Leung and S. Rao, Nucl. Phys. B262 (1985) 517.
 B. Pendleton and G. G. Ross, Phys. Lett. 98B (1981) 291.
 W. Zimmermann, Commun. Math. Phys. 97 (1985) 211;
J. Kubo, K. Sibold and W. Zimmermann, Phys. Lett. B200 (1989) 191	

\bibitem{Bj}Bjorken, J. D. (1968). Current Algebra at Small Distances, in 
Proceedings of the International School of Physics Enrico Fermi 
Course XLI, J. Steinberger, ed., Academic Press, New York, pp. 55-81.

\bibitem{PDF} H.C. Lai et al. ``Global QCD analysis and the CTEQ parton 
distribution '' Phys. Rev. {bf D} Vol. {\bf 51} number 9, p. 4763 (1995);
I used: ``Nuclear CTEQ parton distribution functions'': 
http://projects.hepforge.org/ncteq/   



\bibitem{Fermienergy}
Wikipedia, ``Fermi energy''.

\bibitem{Fewell}
Fewell, M. P. (1995). "The atomic nuclide with the highest mean binding 
energy"American Journal of Physics {\bf 63} (7): 653 - 658. 
Bibcode 1995AmJPh..63..653F 
(http://adsabs.harvard.edu/abs/1995AmJPh..63..653F). doi:10.1119/1.17828 
(http://dx.doi.org/10.1119\%2F1.17828)


\bibitem{beta} M. Fischler and J. Oliensis, Two loop Corrections to 
Beta Functions for the Higgs-Yukawa Coupling Constant'', Physics Letters 
Vol.{\bf 119B}, number 4,5,6; Wikipedia``Top quark'' hop to Yukawa coupling.



\bibitem{Knee}
See figure 2.2 of Sokolsky, P. 1989. Introduction to Ultrahigh 
Energy Cosmic Ray Physics. Redwood City: Addision-Wesley Publishing Company.
See alternatively: 
$``http://imagine.gsfc.nasa.gov/docs/features/topics/snr_group/cr-knee.html''$

\bibitem{Gaisser}
 Gaisser, arXiv:astro-ph/0608553v1 (2006)
\bibitem{explanation}
  Raymond , pp. 683 - 684
Science 7 August 2009:
Vol. 325. no. 5941, pp. 683 - 684
DOI: 10.1126/science.1177743




\bibitem{Tully} Chris Tully, ``Experimental Aspects of Higgs Boson Searches 
at  the LHC'', PiTP, IAS Princeton, Summer 2005.



\bibitem{Higgsprediction}
C. D. Froggatt and H. B. Nielsen, Phys. Lett. {\bf B 368} (1996) 96;

%\cite{Froggatt:2001pa}
%\bibitem{Froggatt:2001pa}
  C.~D.~Froggatt, H.~B.~Nielsen and Y.~Takanishi,
  %``Standard model Higgs boson mass from borderline metastability of the
  %vacuum,''
  Phys.\ Rev.\  D {\bf 64} (2001) 113014
  [arXiv:hep-ph/0104161].
  %%CITATION = PHRVA,D64,113014;%%

\bibitem{boundfirst}
%\cite{Froggatt:2004mv}
%\bibitem{Froggatt:2004mv}
  C.~D.~Froggatt, L.~V.~Laperashvili and H.~B.~Nielsen,
  ``A new bound state 6t + 6anti-t and the fundamental-weak scale 
hierarchy  in
  the standard model,''
  arXiv:hep-ph/0410243.
  %%CITATION = HEP-PH/0410243;%%

\bibitem{bound}
%\cite{Froggatt:2008hc}
%\bibitem{Froggatt:2008hc}
  C.~D.~Froggatt and H.~B.~Nielsen,
  %``Remarkable coincidence for the top Yukawa coupling and an approximately
  %massless bound state,''
  Phys.\ Rev.\  D {\bf 80} (2009) 034033
  [arXiv:0811.2089 [hep-ph]].
  %%CITATION = PHRVA,D80,034033;%%
%\cite{Froggatt:2008uy}
%\bibitem{Froggatt:2008uy}
  C.~D.~Froggatt and H.~B.~Nielsen,
  ``New Bound States of several Top-quarks bound by Higgs Exchange,''
  arXiv:0810.0475 [hep-ph].
  %%CITATION = ARXIV:0810.0475;%%
%\cite{Froggatt:2008ns}
%\bibitem{Froggatt:2008ns}
  C.~D.~Froggatt, L.~V.~Laperashvili, R.~B.~Nevzorov, H.~B.~Nielsen and 
C.~R.~Das,
  ``New Bound States of Top-anti-Top Quarks and $T^-$ balls Production at
  %Colliders (Tevatron, LHC, etc.),''
  arXiv:0804.4506 [hep-ph].
  %%CITATION = ARXIV:0804.4506;%%
%\cite{Froggatt:2005js}
%\bibitem{Froggatt:2005js}
  C.~D.~Froggatt and H.~B.~Nielsen,
  %``Dark matter from encapsulated atoms,''
  arXiv:astro-ph/0512454.
  %%CITATION = ASTRO-PH/0512454;%%
%\cite{Froggatt:2005fk}
%\bibitem{Froggatt:2005fk}
  C.~D.~Froggatt and H.~B.~Nielsen,
  ``Cryptobaryonic dark matter,''
  Phys.\ Rev.\ Lett.\  {\bf 95} (2005) 231301
  [arXiv:astro-ph/0508513].
  %%CITATION = PRLTA,95,231301;%%
%\cite{Froggatt:2004mv}
%\bibitem{Froggatt:2004mv}
%  C.~D.~Froggatt, L.~V.~Laperashvili and H.~B.~Nielsen,
%  %``A new bound state 6t + 6anti-t and the fundamental-weak scale 
%%hierarchy  in
%  %the standard model,''
%  arXiv:hep-ph/0410243.
%  %%CITATION = HEP-PH/0410243;%%

%\cite{Das:2008an}
%\bibitem{Das:2008an}
  C.~R.~Das, C.~D.~Froggatt, L.~V.~Laperashvili and H.~B.~Nielsen,
  %``New Bound States of Heavy Quarks at LHC and Tevatron,''
  arXiv:0812.0828 [hep-ph].
  %%CITATION = ARXIV:0812.0828;%%

%\cite{Das:2009by}
%\bibitem{Das:2009by}
  C.~R.~Das, C.~D.~Froggatt, L.~V.~Laperashvili and H.~B.~Nielsen,
  %``New Bound States of Top and Beauty Quarks at the Tevatron and LHC,''
  arXiv:0908.4514 [hep-ph].
  %%CITATION = ARXIV:0908.4514;%%


\bibitem{hierarchybound}
%\cite{Froggatt:2004bh}
%\bibitem{Froggatt:2004bh}
  C.~D.~Froggatt, H.~B.~Nielsen and L.~V.~Laperashvili,
  %``Hierarchy-problem and a bound state of 6 t and 6 anti-t,''
  Int.\ J.\ Mod.\ Phys.\  A {\bf 20} (2005) 1268
  [arXiv:hep-ph/0406110].
  %%CITATION = IMPAE,A20,1268;%%

\bibitem{Stonybrook}
 M.Yu. Kuchiev, V.V. Flambaum and E. Shuryak, Phys. Rev. D 78, 077502 (2008)
[arXiv:0808.3632].

%\cite{Kuchiev:2010hz}
%\bibitem{Kuchiev:2010hz}
  M.~Y.~.Kuchiev, V.~V.~Flambaum,
  ``Radiative corrections in fermion bags bound by Higgs boson exchange,''
  
  [arXiv:1012.0902 [hep-ph]].

%\cite{Kuchiev:2010ux}
%\bibitem{Kuchiev:2010ux}
  M.~Y.~.Kuchiev,
  %``Multi-fermion states for heavy fermions bound via Higgs exchange,''
  Phys.\ Rev.\  {\bf D82}, 127701 (2010).
  [arXiv:1009.2012 [hep-ph]].



%\cite{Kuchiev:2010ia}
%\bibitem{Kuchiev:2010ia}
  M.~Y.~.Kuchiev,
  %``Amplitudes of radiative corrections in fermion bags bound by Higgs boson exchange,''
  
  [arXiv:1012.0903 [hep-ph]].

%\cite{Crichigno:2010ky}
%\bibitem{Crichigno:2010ky}
  M.~P.~Crichigno, V.~V.~Flambaum, M.~Y.~.Kuchiev, E.~Shuryak,
  %``The W-Z-Top Bags,''
  Phys.\ Rev.\  {\bf D82}, 073018 (2010).
  [arXiv:1006.0645 [hep-ph]].

%\cite{Kuchiev:2008fd}
%\bibitem{Kuchiev:2008fd}
  M.~Y.~.Kuchiev, V.~V.~Flambaum, E.~Shuryak,
  %``On bound states of multiple t-quarks due to Higgs exchange,''
  Phys.\ Rev.\  {\bf D78}, 077502 (2008).
  [arXiv:0808.3632 [hep-ph]].


\bibitem{MPP}
%THREE IMPORTANT MPP PAPERS:
%\bibitem{MPPfirst}
%\cite{Bennett:1993pj}
%\bibitem{Bennett:1993pj}
  D.~L.~Bennett and H.~B.~Nielsen,
  ``Predictions for nonAbelian fine structure constants from
  multicriticality,''
  Int.\ J.\ Mod.\ Phys.\  A {\bf 9} (1994) 5155
  [arXiv:hep-ph/9311321].
  (CITATION = IMPAE,A9,5155)

%\bibitem{MPPsecth}
%\cite{Bennett:1996hx}
%\bibitem{Bennett:1996hx}
  D.~L.~Bennett,
  ``Multiple point criticality, nonlocality and fine tuning in fundamental
  physics: Predictions for gauge coupling constants gives alpha**(-1) =  136.8
  +- 9,''
  arXiv:hep-ph/9607341.
  (CITATION = HEP-PH/9607341)

%\bibitem{MPPthirdel}
%\cite{Bennett:1996vy}
%\bibitem{Bennett:1996vy}
  D.~L.~Bennett and H.~B.~Nielsen,
  ``Gauge couplings calculated from multiple point criticality yield
  alpha**(-1) = 137+-9: At last, the elusive case of U(1),''
  Int.\ J.\ Mod.\ Phys.\  A {\bf 14} (1999) 3313
  [arXiv:hep-ph/9607278].
  (CITATION = IMPAE,A14,3313);

\bibitem{SSC}
 Michelle Mittelstadt (AP) (October 22, 1993). "Congress officially kills 
collider project". Sun Journal (Lewiston). p. 7. 
``http://news.google.com/newspapers?id=kGAgAAAAIBAJ\&sjid=umUFAAAAIBAJ\&dq
=cost\%20overrun\%20ssc\&pg=3808%2C4981568.'' 
Retrieved 2010-06-28.

Cramer, John G. (May 1997). "The Decline and Fall of the SSC". 
The Alternate View column. Analog Science Fiction and Fact Magazine. 
Archived from the original on 2001-02-23. 
http://web.archive.org/web/20010223042619/
http://www.npl.washington.edu/av/altvw84.html. 
Retrieved 2010-07-11. 


\end{thebibliography}
\end{document}